# Spin-glass state induced by Mn-doping into a moderate gap layered semiconductor SnSe$_2$

Hongrui Huang[*1], Azizur Rahman[2], Jianlin Wang[3, 4], Yalin Lu[3, 4], Ryota Akiyama[†1], and Shuji Hasegawa[1]

[1]*Department of physics, The University of Tokyo, 7-3-1 Hongo, Bunkyo-ku, Tokyo, Japan*

[2]*Department of Physics, University of Science and Technology of China, Hefei 230026, China*

[3]*Hefei National Laboratory for Physical Sciences at the Microscale, University of Science and Technology of China, Hefei 230026, China*

[4]*Anhui Laboratory of Advanced Photon Sciences and Technology, University of Science and Technology of China, Hefei 230026, China*

**Abstract**

Various types of magnetism can appear in emerging quantum materials such as van der Waals layered ones. Here, we report the successful doping of manganese atoms into a post-transition metal dichalcogenide semiconductor: SnSe$_2$. We synthesized a single crystal Sn$_{1-x}$Mn$_x$Se$_2$ with $x = 0.04$ by the chemical vapor transport (CVT) method and characterized it by x-ray diffraction (XRD) and energy-dispersive x-ray spectroscopy (EDS). The magnetic properties indicated a competition between coexisting ferromagnetic and antiferromagnetic interactions, from the temperature dependence of the magnetization, together with magnetic hysteresis loops. This means that magnetic clusters having ferromagnetic interaction within a cluster form and the short-range antiferromagnetic interaction works between the clusters; a spin-glass state appears below ~ 60 K. Furthermore, we confirmed by *ab initio* calculations that the ferromagnetic interaction comes from the 3*d* electrons of the manganese dopant. Our results offer a new material platform to understand and utilize the magnetism in the van der Waals layered materials.

**Introduction**

Two-dimensional (2D) van der Waals (vdW) layered materials with a chemical formula of MX$_2$ (M = metal, X = S, Se) have been received great attention in recent years due to the ease of processing and their novel electronic properties [1]-[5]. Tin diselenide (SnSe$_2$), a post-transition metal dichalcogenide, is a member of the 2D layered materials and has revealed rich physics, such as the charge density wave [6] and the gate-induced superconductivity [7]. In addition, researches on its potential optoelectronic and thermoelectric applications have been reported recently [8][9]. Besides, for realizing spin dependent transport, the ferromagnetic properties can be induced into the SnSe$_2$ system by doping transitional metal atoms or nonmetal atoms, resulting in a diluted magnetic semiconductor [10][11], which will be of importance for abundant functional spintronic devices.

[*] huanghr@surface.phys.s.u-tokyo.ac.jp

[†] akiyama@surface.phys.s.u-tokyo.ac.jp

On the other hand, the spin-glass state, in which the short-range magnetic order is introduced without the long-range magnetic order, has also attracted much interest for its theoretical models and simulation methods [12]. For the competition between the ferromagnetic state and the spin-glass state, previous researches have mostly been done in dilute metallic alloys [15] and some perovskite oxides [16]. Several models, such as the Sherrington-Kirkpatrick model and the short-range Edwards-Anderson model have been established in order to explain the transition from the short-range magnetic order to the long-range one [12]. However, most of the systems above are three-dimensional (3D) systems, while there has been little work on the study of 2D layered materials having ferromagnetic and spin-glass states coexisting.

In this work, we have synthesized a crystalline Mn-doped SnSe$_2$ crystal by the chemical vapor transport (CVT) method for the first time, and investigated the magnetic properties of Sn$_{1-x}$Mn$_x$Se$_2$ with $x = 0.04$ by measuring the temperature and magnetic field dependences of magnetization as well as the structural analyses. We have observed the coexistence of ferromagnetic (FM), antiferromagnetic (AFM), and spin-glass states at low temperature in the system. Furthermore, by the density functional theory (DFT) calculation, we have found the appearance of a spin-polarized state around the bottom of the conduction band by Mn-doping, which is suggested to be the origin of magnetism in Sn$_{0.96}$Mn$_{0.04}$Se$_2$ crystal.

**Method**

First, the single crystal SnSe$_2$ flakes were prepared by the CVT method [13]. Stoichiometric amounts of high-purity (>99.99%) Sn and Se powder were mixed and sealed into a quartz ampoule with 200 mm in length, 10 mm in diameter, and 1.5 mm in the wall thickness. The ampoule was put into a dual zone tube furnace, and the temperatures at the hot and the cold ends of the ampoule were kept at 660°C and 500°C, respectively, for 72 hours. Then, it was slowly cooled down to room temperature (RT), and some flakes (2 mm × 3 mm - 7 mm × 10 mm) of the single crystalline SnSe$_2$ were obtained at the cold end.

Next, the single crystal Sn$_{1-x}$Mn$_x$Se$_2$ with $x = 0.04$ was prepared by the two-step method. First, stoichiometric amounts of high-purity (> 99.99%) Sn, Mn and Se powder were mixed and sealed into a quartz ampoule. The ampoule was heated at 600 °C and kept for 48 hours to ensure that the solid phase reaction among the three materials proceeded. Then, the intermediate product was grinded and put into another ampoule for the CVT method. The growth conditions, such as the temperature and the growth time, were the same as those used for synthesizing the single crystal SnSe$_2$. Similarly, the flakes of Sn$_{0.96}$Mn$_{0.04}$Se$_2$ (2 mm × 2 mm - 5 mm × 7 mm) were acquired at the cold end after the reaction.

**Characterization**

Figure 1 (a) shows the crystal structure of a unit layer (UL) of the hexagonal $SnSe_2$, which has a layered van der Waals (vdW) type structure with an interlayer thickness of 0.62 nm. Within a UL, Sn atoms are sandwiched between two layers of Se atoms, and the lateral distance between the nearest Sn atoms is 0.38 nm. Figure 1 (b) shows a structural model of Mn-doped $SnSe_2$ layer assumed for the DFT calculation; the vertex Sn atoms in the 5×5 supercell of the $SnSe_2$ UL are replaced by Mn atoms because of the Mn concentration of 0.04 at. % .

Figures 1 (c) and (d) show the room-temperature $2\theta$-$\theta$ scans of the $SnSe_2$ and $Sn_{0.96}Mn_{0.04}Se_2$, respectively, by x-ray diffraction (XRD) with Rigaku Smart Lab. Because of the vdW layered structure of the single crystal flakes, the strongest diffraction peaks correspond to the layer spacing in *c*-axis. The Miller indices of the diffraction peaks are shown in the figures, and from the calculation, the distance between two adjacent ULs is increased by ~ 0.19 % by Mn-doping.

Besides, the stoichiometric ratio of samples was confirmed by the energy-dispersive X-ray spectroscopy (EDS) measurements as shown in Figures 1 (e) and (f) for $SnSe_2$ and $Sn_{0.96}Mn_{0.04}Se_2$, respectively. The observed spectra indicate that the stoichiometric ratios for both flakes are the same as the aimed composition within the error.

**Magnetic properties**

The temperature dependences of the magnetization (*M-T* curves) for $SnSe_2$ and $Sn_{0.96}Mn_{0.04}Se_2$ were measured. To compare the magnetic properties of the samples with and without Mn-doping, the samples were firstly cooled down to 2 K under zero magnetic field, and then after the magnetic field of 10 kOe perpendicular to the sample surface was applied, the samples were warmed up to the room temperature with measuring magnetization (ZFC process). After temperature reaches the room temperature, the magnetic field of 10 kOe perpendicular to the sample surface was applied, and then the samples were cooled down to 2 K again with measuring magnetization (FC process). All measurements were done with the Quantum Design MPMS 3 Superconducting Quantum Interference Device (SQUID) magnetometer system.

As shown in Fig. 2 (a), the $SnSe_2$ sample (red symbols) shows quite small magnetizations in both the ZFC and FC cases; they show actually negative magnetization, diamagnetic with negligible temperature dependence. For the $Sn_{0.96}Mn_{0.04}Se_2$ sample (blue symbols), on the contrary, the magnetization *M* increases slowly as the temperature decreases down to $T_C$ ~ 66 K; below $T_C$ it suddenly increases and shows a clear bifurcation between the FC and ZFC processes at $T_B$= 62 K. For the FC curve, it increases monotonically with a nearly constant slope as the temperature decreases, while for the ZFC curve it first increases with a gradual slope with further cooling and at $T_F$ ~ 12 K it turns to decrease.

To understand the magnetic properties of the $Sn_{0.96}Mn_{0.04}Se_2$ sample more, we made a 1/*M-T* plot as shown in Fig. 2(b). For both ZFC and the FC curves, they obey the Curie-Weiss (C-W) law fitted by a dotted straight line. As a result, we estimated the Curie temperature to be $T_C = 66$ K where the $1/M$-$T$ curve starts to deviate from the fitted line. The Weiss temperature $T_W$ was also estimated by the $T$-intercept of the fitted line to be -170 K with the formula $\chi = \frac{C}{T-T_W}$, where $\chi$ is the magnetic susceptibility, $C$ is the Curie constant, and $T$ is the system temperature. The negative $T_W$ indicates that this system shows AFM interaction.

However, as mentioned below, by measuring the magnetic field dependence of magnetization, we have found the magnetic hysteresis below $T_C$, which is different from the behavior of a simple AFM system. In Figure 3, the magnetic field dependences of magnetization (*M-H* curves) for a flake of $Sn_{0.96}Mn_{0.04}Se_2$ at $T$ = 3, 9, 18, 60, and 70 K are shown. At 70 K which is above $T_C$, there is no magnetic hysteresis, meaning a paramagnetic behavior. As shown in the inset of Fig. 3, with decreasing temperature, at ~ 60 K which is below $T_C$, a magnetic hysteresis loop having residual magnetization starts to be seen, meaning a FM order. The magnitude of residual magnetization and the coercivity increase as the temperature decreases, and the largest coercivity is 4.25 kOe at 3 K. All hysteresis loops show the unsaturated component even at 50 kOe, which means that the sample contains paramagnetic Mn atoms which do not participate in the FM order.

Considering results shown in Fig. 2, a spin-glass state is indicated by characteristic behaviors such as the separating curves between ZFC and FC samples below $T_B$ and the rapid decrease of magnetization below $T_F$ in the ZFC curve [14]. A schematic picture of the temperature dependence of magnetic behaviors in this system is depicted in Fig. 4. Because Mn atoms are randomly doped in the crystal, the magnetic clusters which contain some number of Mn atoms appear as shown by dashed circles. We can assume a FM interaction among Mn spins in short range to make each cluster FM (shown by arrows in Fig. 4). This is because if the AFM interaction works among the Mn spins in short range, the ferromagnetic hysteresis shown in Fig. 3 would not be observed. Furthermore, it is suggested that an AFM interaction works among the clusters because the magnetization goes to zero at the lowest temperature in the ZFC process, resulting in a spin-glass state (freezing of spins). Since there is no external magnetic field during cooling the sample at the ZFC process, the inter-cluster AFM interaction remain, resulting in random orientation of magnetization of each cluster. During warming the sample under the measurement magnetic field for SQUID, directions of magnetization within clusters are gradually aligned along the magnetic field with the help of thermal energy, so that the total magnetization increases by warming in $T<T_F$ region. In $T_F<T<T_B$ region, the thermal fluctuation weakened the magnetization in each cluster, resulting in a decrease of the total magnetization with increasing temperature.

In the FC process, on the other hand, the external magnetic field overcomes the AFM inter-cluster interaction, so that the magnetization of each cluster is aligned to the direction of the magnetic field, resulting in the highest magnetization at the lowest temperature. The magnetization just decreases by warming in $T<T_B$ region by thermal fluctuation. With comparing the degree of alignment of magnetization of each cluster along the magnetic field between the ZFC process and FC process, the FC process should have better degree of alignment because of the initial FM state at the lowest temperature, so that the FC curve shows higher magnetization than the ZFC curve in $T_F<T<T_B$ region. In $T_B \leq T \leq T_C$ region, the thermal fluctuation is strong enough to make the magnetization direction of each cluster in the ZFC process aligns along the magnetic field as good as in the FC process, so that the ZFC and FC curves merge each other there.

In this way, we can understand the magnetic behaviors of the Mn-doped SnSe$_2$ flake by assuming the inner-cluster FM interaction and the inter-cluster AFM interaction. The inner-cluster FM interaction makes magnetic clusters below $T_C$, while the inter-cluster AFM interaction dominantly develops below ~ $T_F$. The spin glass state with magnetic hysteresis is formed due to the competition between the FM and AFM interaction.

In order to understand the electronic origin of the observed magnetic behaviors induced by Mn atoms, we performed the *ab-initio* calculation for both SnSe$_2$ and Sn$_{0.96}$Mn$_{0.04}$Se$_2$ crystals using density functional theory with the Perdew–Burke–Ernzerhof (PBE) scheme for the electronic exchange–correlation functional by the Quantum Espresso package [18] [19]. For the SnSe$_2$ calculation, we adopted the cutoff energy 50 Ry and the 11×11×7 k-point meshes for integrating the Brillouin zones. For the Sn$_{0.96}$Mn$_{0.04}$Se$_2$, the 5×5×1 superlattice was used (Fig. 1(b)) and the k-point meshes were set as 3×3×7. We incorporate the degree of the spin polarization in the calculation for Sn$_{0.96}$Mn$_{0.04}$Se$_2$.

We first compare the density of states (DOS) near the Fermi level of these two material systems as shown in Fig. 5 (a). For the non-doped SnSe$_2$ crystal, there is a moderate band gap stretching across the Fermi level (~ 0.52 eV wide), which is consistent with the previous calculation [20]. With the Mn-doping, however, a new energy level appears near the bottom of conduction band, at ~ 0.45 eV, and the width of the band gap reduces to ~ 0.38 eV. As shown in Fig. 5 (b), when we compare the spin-resolved DOS between the up-spin and the down-spin for the Mn-doped system, the spin polarization of the new energy state is clearly seen. The difference of DOS between up and down spin at ~ 0.45 eV induced by Mn-doping illustrates the origin of the magnetic properties. The projected DOS for analyzing the orbital contribution is shown in Figs. 5 (c) and (d). As is known, the electron orbital configuration of the Mn atom is represented as [Ar] $4s^2 3d^5$. Therefore, we only need to compare the contribution of valence electrons, that is, the electrons in the $4s$ and the $3d$ orbitals. We can see that both the $4s$ and $3d$ electrons show the spin-polarization. Notably, the components of the $3d$ electrons are much stronger than that of the $4s$ electrons, and the shape of the spin-polarized DOS for $3d$ electrons is almost the same as the total spin-resolved DOS shown in the upper figure of Fig. 4(b) around the bottom of the conduction band. Thus, we conclude that the observed magnetic properties stem from the $3d$ electrons of Mn dopants.

## **Conclusion**

We synthesized single-crystal samples of a moderate gap semiconductor SnSe$_2$ and a Mn-doped crystal, Sn$_{0.96}$Mn$_{0.04}$Se$_2$, by the CVT method. The measurements of magnetic properties indicate that Mn dopants induce the FM order ($T_C$ = 66 K), together with the spin-glass state by the mixing contribution of FM and AFM interactions at low temperature below $T_F$ ~ 12 K. The DFT calculation shows that a spin-polarized band appears by Mn-doping around the bottom of the conduction band and suggests that from the analysis of the contribution of orbitals the $3d$ electrons of the Mn atom play a key role for the magnetic properties. The complicated magnetic states in the layered material SnSe$_2$ we have shown here provide an interesting platform for study of low-dimensional magnetism as well as possible applications in spintronics with diluted magnetic semiconductors.

**Acknowledgement**

We thank Zhibo Zhao, Qingmei Wu and Jiameng Cui for their help of measurements on SQUID and XRD systems. The computation in this work has been done using the facilities of the Supercomputer Center, the Institute for Solid State Physics, the University of Tokyo. This work was partly supported by the KAKENHI (20H02616, 20H00342) by JSPS and Natural Science Foundation of China (12004366). HH was supported by Advanced Leading Graduate Course for Photon Science (ALPS).

**Reference**

[1]  Xi X *et al. Nat. Nanotechnol.* **10**, 765–769 (2015).

[2]  Ugeda M. M *et al. Nat. Phys.* **12**, 92–97 (2016).

[3]  Guillamón I *et al. Phys. Rev. Lett.* **101**, 166407 (2008).

[4]  Radisavljevic B *et al. Nat. Nanotechnol.* **6** 147–150 (2011)

[5]  Wang Z *et al. Phys. Rev. Lett.* **117** 056805 (2016)

[6]  Shao Z *et al. Nano Lett.* **19** (8), 5304–5312 (2019)

[7]  J Zeng *et al. Nano Lett.* **18** (2) 1410–1415 (2018)

[8]  D. Martínez-Escobar *et al. Thin Solid Films* **535**, 390 (2013)

[9]  Wu Y *et al. Mater. Today Phys.* **3**, 127-136. (2017)

[10] Luo J *et al. J. Magn. Magn. Mater.* **486**, 165269 (2019)

[11] Dong S *et al. APL Mater.* **4**, 032601 (2016)

[12] Binder K *et al. Rev. Mod. Phys.* **58**, 801. (1986)

[13] Schmidt P *et al.* Rijeka, Croatia: InTech (2013)

[14] Joy P A *et al. J. Phys. Condens. Matter.* **10**, 4811049 (1998)

[15] Campbell I A *et al. Phys. Rev. Lett.* **50**, 201615 (1983)

[16] Yoo Y J *et al. J. Appl. Phys*. **112**, 013903 (2012)

[17] J. D. Browne *et al. Phys. Status Solidi A* **49**, K177 (1978)

[18] M. Ernzerhof *et al. J. Chem. Phys.* 110, 5029 (1999)

[19] Giannozzi P *et al. J. Phys. Condens. Matter.* **21**, 395502 (2009)

[20] R. H Williams *et al. J. Phys. Condens. Matter.* **6** *3631* (1973)

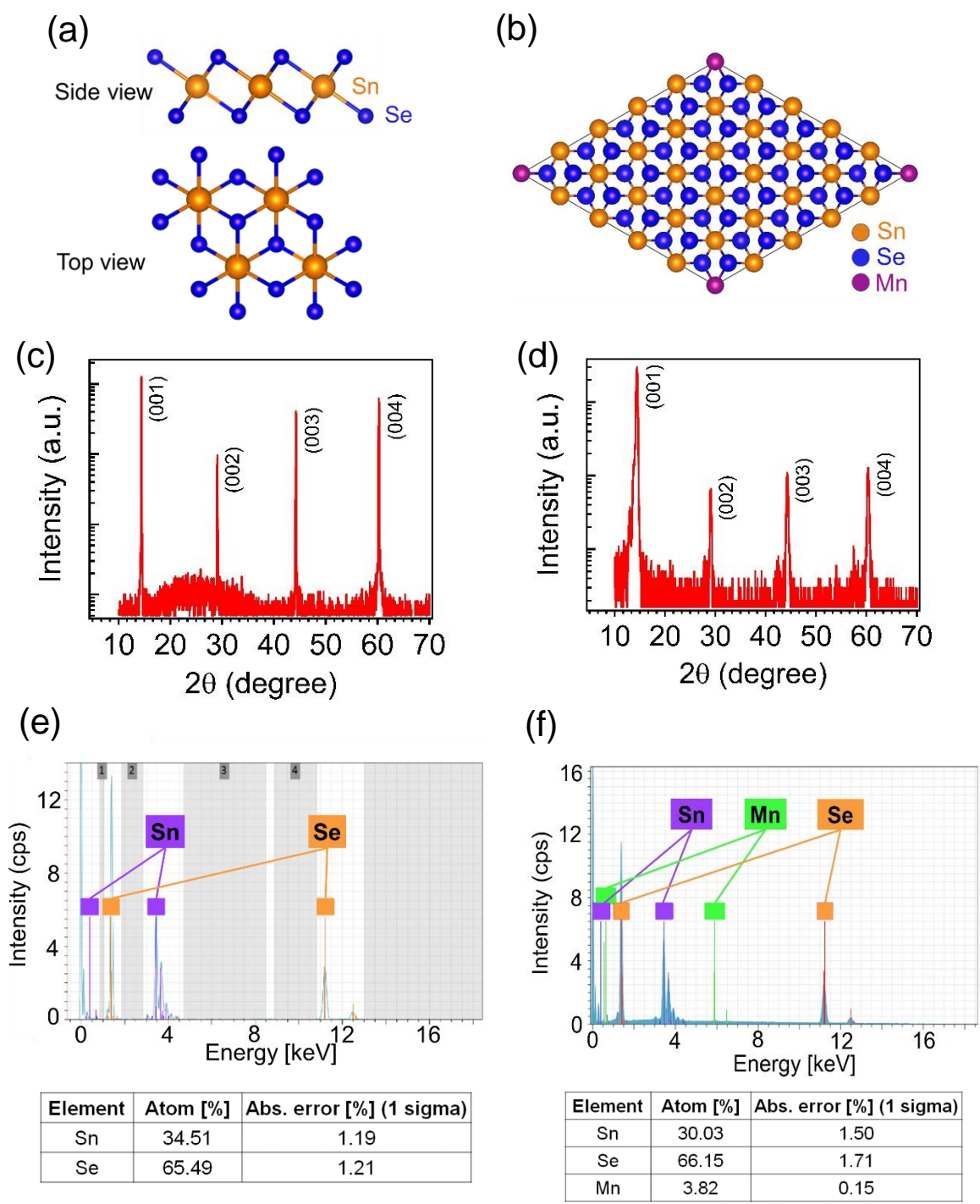

Figure 1. (a) Side view and top view of the atomic structure of a UL of $SnSe_2$. (b) A structural model of the supercell $Sn_{0.96}Mn_{0.04}Se_2$ for the DFT calculation. (c)(d) $2\theta$-$\theta$ scans in XRD for $SnSe_2$ and $Sn_{0.96}Mn_{0.04}Se_2$, respectively. The Miller indices are marked on the corresponding peaks. (e)(f) EDS X-ray analyses for $SnSe_2$ and $Sn_{0.96}Mn_{0.04}Se_2$, respectively. The stoichiometric ratios of each element are shown on the tables below.

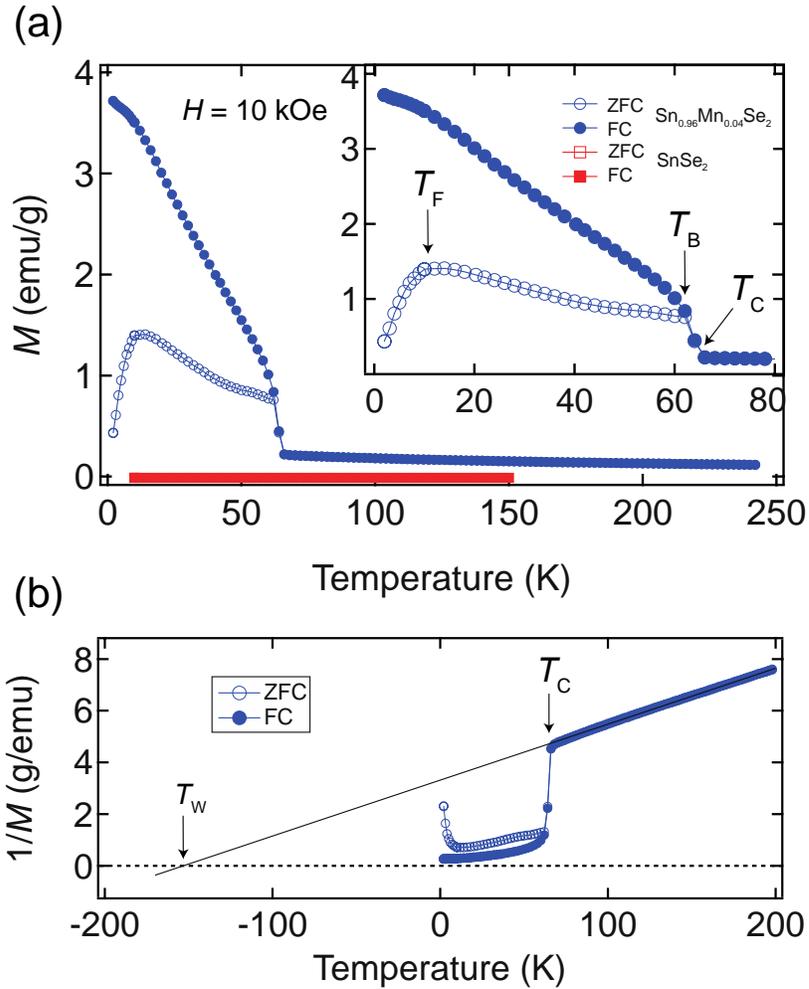

Figure 2. (a) Temperature dependence of magnetization for $SnSe_2$ (red rectangles) and $Sn_{0.96}Mn_{0.04}Se_2$ (blue circles). The FC (ZFC) curves are labelled by the open (filled) symbols. For $SnSe_2$, magnetization is diamagnetic and very small comparing with that of $Sn_{0.96}Mn_{0.04}Se_2$. For the $Sn_{0.96}Mn_{0.04}Se_2$, magnetization starts to increase with decreasing temperature at $T_C$ (66 K), and the FC and ZFC curves are separated below $T_B$ (62 K). With cooling more, the ZFC curves shows a peak $T_F$ (~ 12 K) whereas the FC curve increases monotonically. (b) The Curie-Weiss fitting for the $1/M$-$T$ curves of $Sn_{0.96}Mn_{0.04}Se_2$. The Weiss temperature is estimated as the $T$-intercept of the fitted line (-170 K).

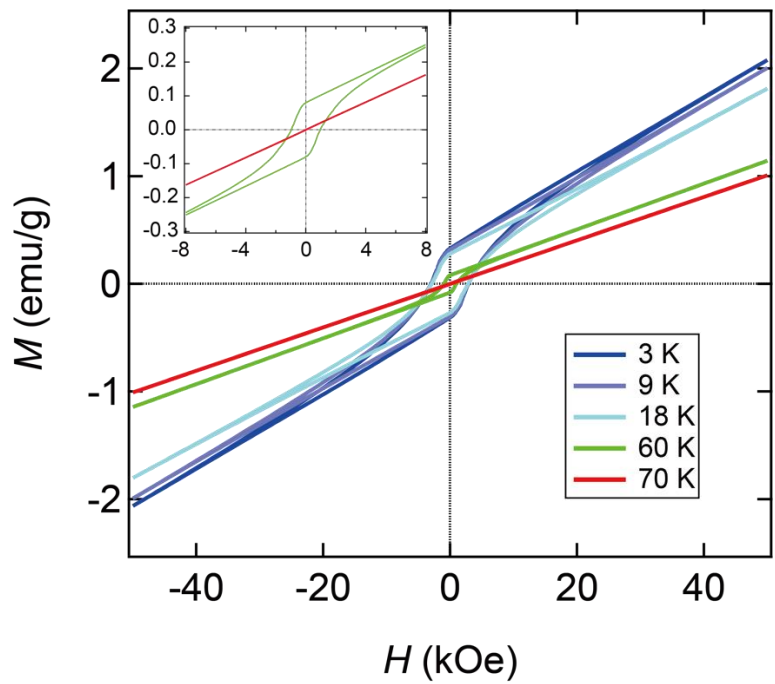

Figure 3. Magnetic field dependance of magnetization for $Sn_{0.96}Mn_{0.04}Se_2$ at different temperatures. Inset: curves below 60 K show hysteresis loops meaning the ferromagnetic component whereas above 70 K the curve is paramagnetic.

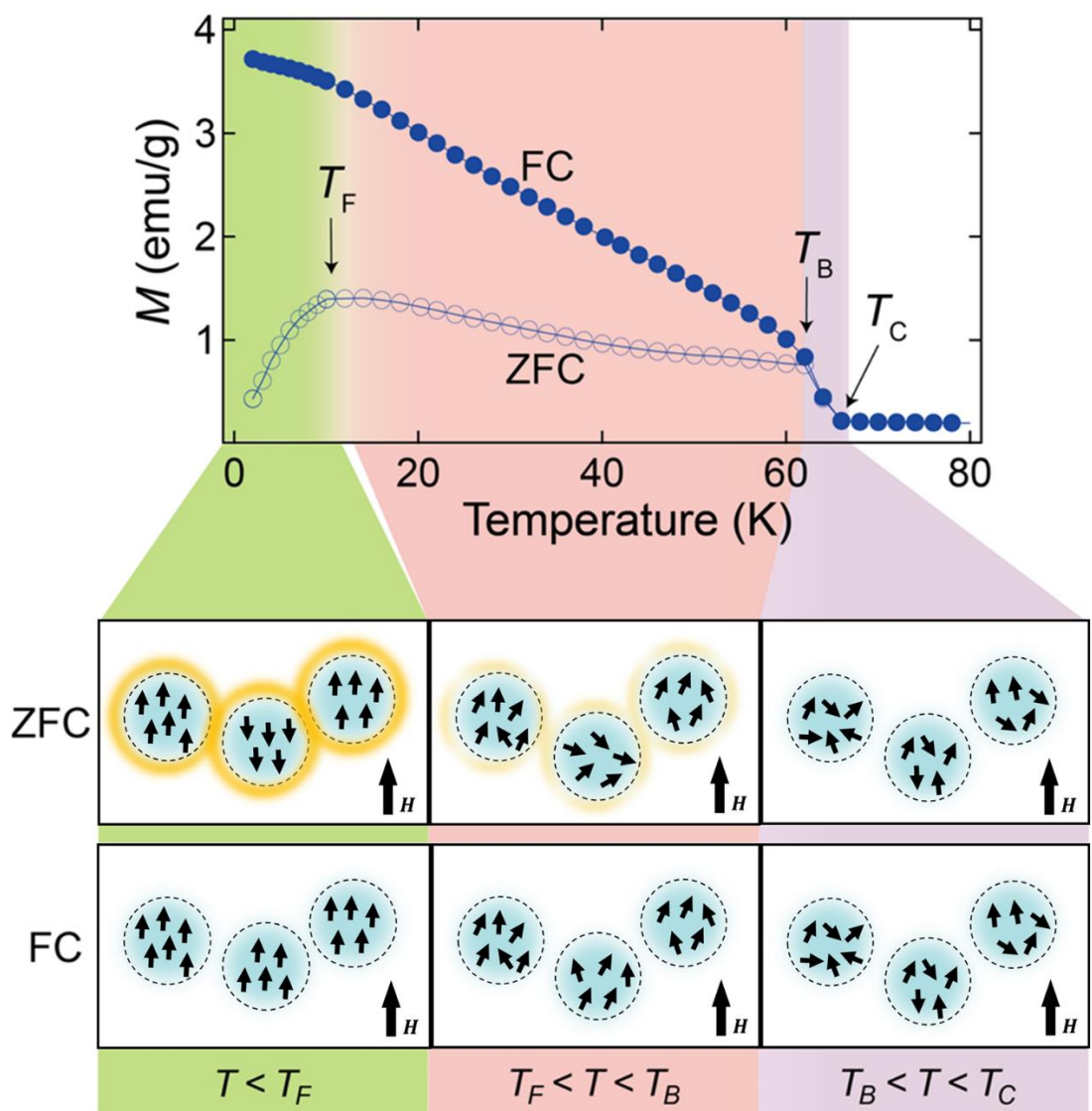

Figure 4. The schematic diagram of departure dependences of magnetization at each temperature region for both the FC and ZFC process. The black arrows represent the local magnetic moment of the Mn atoms. The dashed circles represent the magnetic clusters. Light blue shadows indicate the inner-cluster FM interaction, while the yellow regions are the inter-cluster AFM interaction. When the temperature increases, the direction of the magnetic moment in a cluster becomes more fluctuated, which weakens the inner-cluster FM interaction.

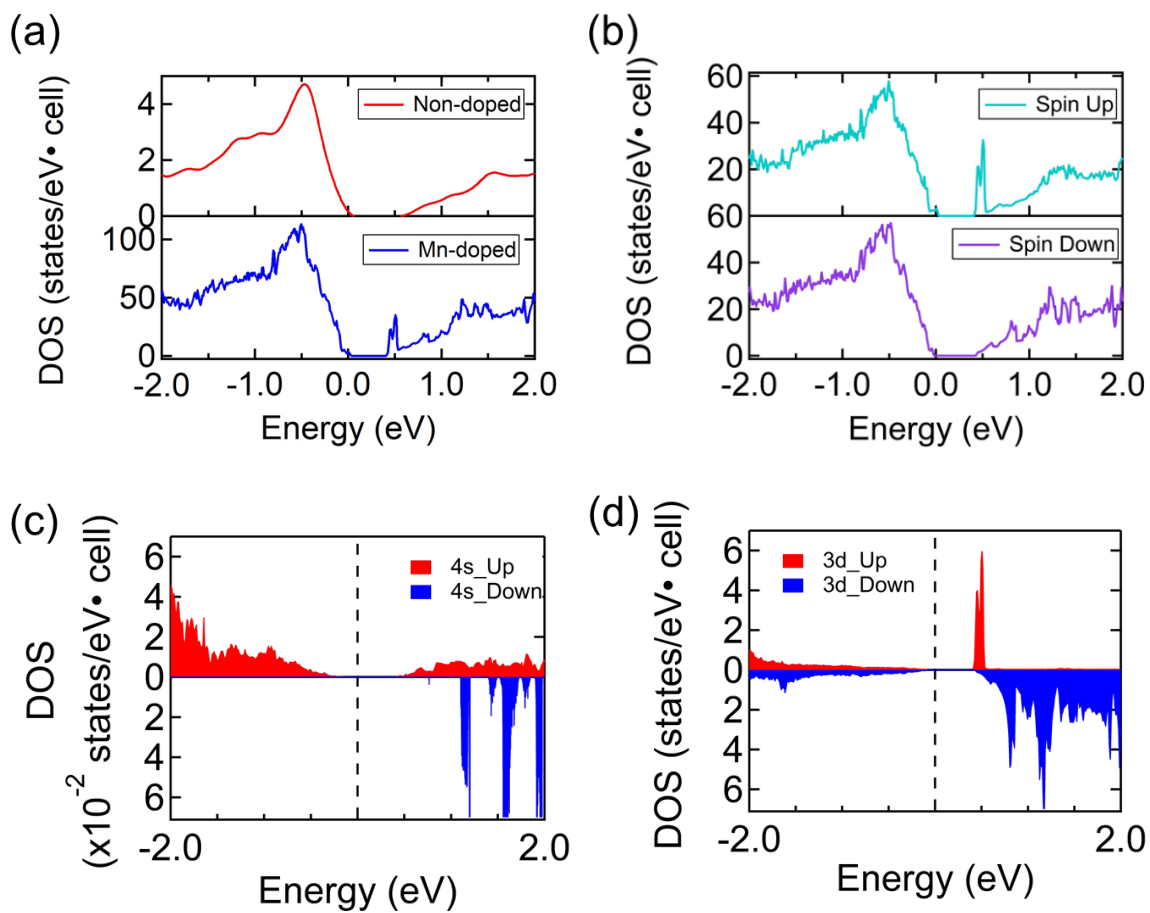

Figure 5. (a) The total DOS of the SnSe$_2$ in comparison to the Sn$_{0.96}$Mn$_{0.04}$Se$_2$. (b) Spin-resolved DOS of the Sn$_{0.96}$Mn$_{0.04}$Se$_2$. (c), (d) Projected spin-resolved DOS of the Mn atoms for the 4s orbital and 3d orbital, respectively. The dash line illustrates the Fermi level at zero eV.